\documentclass[11pt,a4in]{article}
\usepackage{graphicx}
\usepackage[utf8]{inputenc}
\usepackage[T1]{fontenc}
\usepackage{geometry}
\usepackage[francais,english]{babel}
\begin{document}
\clearpage
\pagenumbering{gobble}
\thispagestyle{empty}
\renewcommand{\thesection}{\Roman{section}}
\begin{center}
\large{\bf Few-body model approach to the lowest bound S-state of non-symmetric exotic few-body systems}\\
\end{center}
\vspace*{0.5cm}
\hspace*{2cm}\normalsize{\bf Md. A. Khan* and M. Hasan}\\
\footnotesize
\hspace*{2cm}Department of Physics, Aliah University, \\ \hspace*{2cm} IIA/27, Newtown, Kolkata-700156, India\\ 
\hspace*{2cm}{\it Email:} $ drakhan.rsm.phys@gmail.com; drakhan.phys@aliah.ac.in$\\
\vspace{0.5cm}
\rm
\begin{abstract}
Lowest bound S-state energy of Coulomb three-body systems ($N^{Z+}\mu^-e^-$) having a positively charged nucleus of charge number Z ($N^{Z+}$), a negatively charged muon ($\mu^-$) and an electron ($e^-$), is investigated in the framework of hyperspherical harmonics expansion method. A Yukawa type Coulomb potential with an adjustable screening parameter ($\lambda$) is chosen for the 2-body subsystems. In the resulting Schr\"{o}dinger equation (SE), the three-body relative wave function is expanded in the complete set of hyperspherical harmonics (HH). Thereafter use of orthonormality of HH in the SE, led to a set of coupled differential equations which are solved numerically to get the energy (E) of the systems investigated.
\end{abstract}
\hspace{1.0cm}{\it Keywords:} Hyperspherical Harmonics (HH), Raynal-Revai Coefficient (RRC), \\\hspace*{2.5cm}Renormalized Numerov Method (RNM), Exotic Ions.\\ \hspace*{1.0cm}{\it PACS:} 02.70.-c, 31.15.-Ar, 
31.15.{\cal J}a, 36.10.{\cal E}e.

\section{Introduction}
Helium like exotic systems may those can be formed by replacing an orbital electron in atoms (or ions) by an exotic particle of the same electric charge can in general be represented by $N^{Z+}x^-e^-$, where N represents the atomic nucleus having Z number of protons and x represents exotic particles like muon, kaon, taon, baryon or their anti-matters. Among the many possible exotic particles, the one which is the most fascinating is the muon, which is sometimes called as big cousin of electron. Muons are widely used as an important probe to the physics beyond standard model. This is because this particle has fair chances of interacting with virtual particles as inferred by the anomalous value of its measured magnetic moment over theoretical predictions [1]. literature survey reveals the fact, studies of muon decay may yield very important information on the overall strength and chiral structure of weak interactions in addition to charged-lepton-flavor-violating processes [2]. Spectroscopy of muonium and muonic atoms gives unmatched determinations of fundamental quantities including the magnetic moment ratio $\mu_{\mu}/\mu_p$, lepton mass ratio $m_{\mu}/m_e$, and proton charge radius $r_p$ [2]. Moreover, muon capture experiments are exploring elusive features of weak interactions involving nucleons and nuclei. For production of muon one may refer to decay of pions produced during bombardment of target by fast moving proton beams apart from its cosmic origin. Out of several types of exotic few-body systems, the one, which contains at least one muon are the most exciting. These exciting muonic atoms/ions can be formed in exact analogy with electronic atoms/ions. It is interesting to note that muon being nearly 200 times heavier than an electron revolves round the nucleus in orbit much smaller than that of an orbital electron. In fact, a 1S muon spends about one half of its periodic life inside the nucleus in muonic atoms/ions having nuclear charge number $Z\sim 82$ [3]. It is worth mentioning here that the spectra of muonic atom (for Z$\sim$ 82) lies in the X-ray region. Further, muonic atoms/ions ($N^{Z+}\mu^-e^-$) are the rare kind of atomic three-body systems having no restriction due to Pauli principle for the muon and electron being non-identical fermions. These atoms/ions are generally formed as by-products of the process of muon catalyzed fusion, hence are useful to understand the fusion reactions in a proper way [4-5]. Although, hyperfine structure of these atoms are studied by several groups to interpret the nature of electromagnetic interaction between the electron and negatively charged muon [6-7], less attention have been given towards {\it ab-initio} calculation for the bound state observables of such exotic atoms/ions in the few-body model approach.

As exotic particles are mostly unstable, their parent atoms (or ions) are also very short lived, and can be formed by trapping the accelerated exotic particle(s) inside matter at the cost of same number of electron(s). The trapped exotic particle revolves round the nucleus of the target atom in orbit of radius equal to that of the electron before its ejection from the atom. Which subsequently cascades down the ladder of resulting exotic atomic states by the  emission of X-rays and Auger transitions before being lost on its way to the atomic nucleus. If the absorbed exotic particle is a negatively charged muon, it passes through various intermediate environments undergoing scattering from atom to atom (in a way similar to that for a free electron) resulting in a gradual loss of its energy until it is captured into an atomic orbit. In the lowest energy level (1S), it feels only Coulomb interaction with nuclear protons and also experiences weak interaction with the rest of the nucleons.

 In this work, we considered only those atoms/ions in which the positively charged nucleus is being orbited by one electron and one negatively charged muon. We adopt hyperspherical harmonics expansion (HHE) approach for a systematic study of the ground state of atoms/ ions of the form $N^{Z+}\mu^-e^-$ assuming a three-body model for each of them. In our model, we assume that the electromagnetic interaction of the valence particles with the nucleus is weak enough to influence the internal structure of the nucleus. Again, the fact that the muon is much lighter than the nucleus allows us to regard the nucleus to remain a static source of electrostatic potential. However, a hydrogen-like two-body model, consisting a quasi-nucleus ($\mu^-N^{(Z-2)+}$, formed by the muon and the nucleus plus an orbital electron will also be tested due to significantly smaller size of muonic orbital compared to an electronic orbital.

In HHE approach to a general three-body system consisting three unequal mass particles the choice of Jacobi coordinates, corresponds to three different partitions and in the $i^{th}$ partition, particle labeled by $\lq i$' acts as a spectator while the remaining two labeled $\lq j$' and $\lq k$' form the interaction pair. So the total potential is the sum of three two-body potential terms (i.e. $V = V_{jk}(r_{jk}) + V_{ki}(r_{ki}) + V_{ij}(r_{ij})$). For the computation of matrix element of V($r_{ij}$), potential of the $(ij)$ pair, the chosen HH is expanded in the set of HH corresponding to the partition in which $\vec{r_{ij}}$ is proportional to the first Jacobi vector [8] by the use of Raynal-Revai coefficients (RRC) [9]. The binding energies obtained for the lowest bound s-state are compared with the ones of the literature wherever available . 

In Section II, we briefly introduce the hyperspherical coordinates and the scheme of the transformation of HH belonging  to two different partitions. Results of calculation and discussions will be presented in Section III and finally we shall draw our conclusion in section IV.

\section{HHE Method}
The choice of Jacobi coordinates for systems of three particles of mass $m_{i}$, $m_{j}$, $m_{k}$ is shown in Figure 1.
\begin{figure}
\centering
\fbox{\includegraphics[width=0.75\linewidth, height=0.35\linewidth]{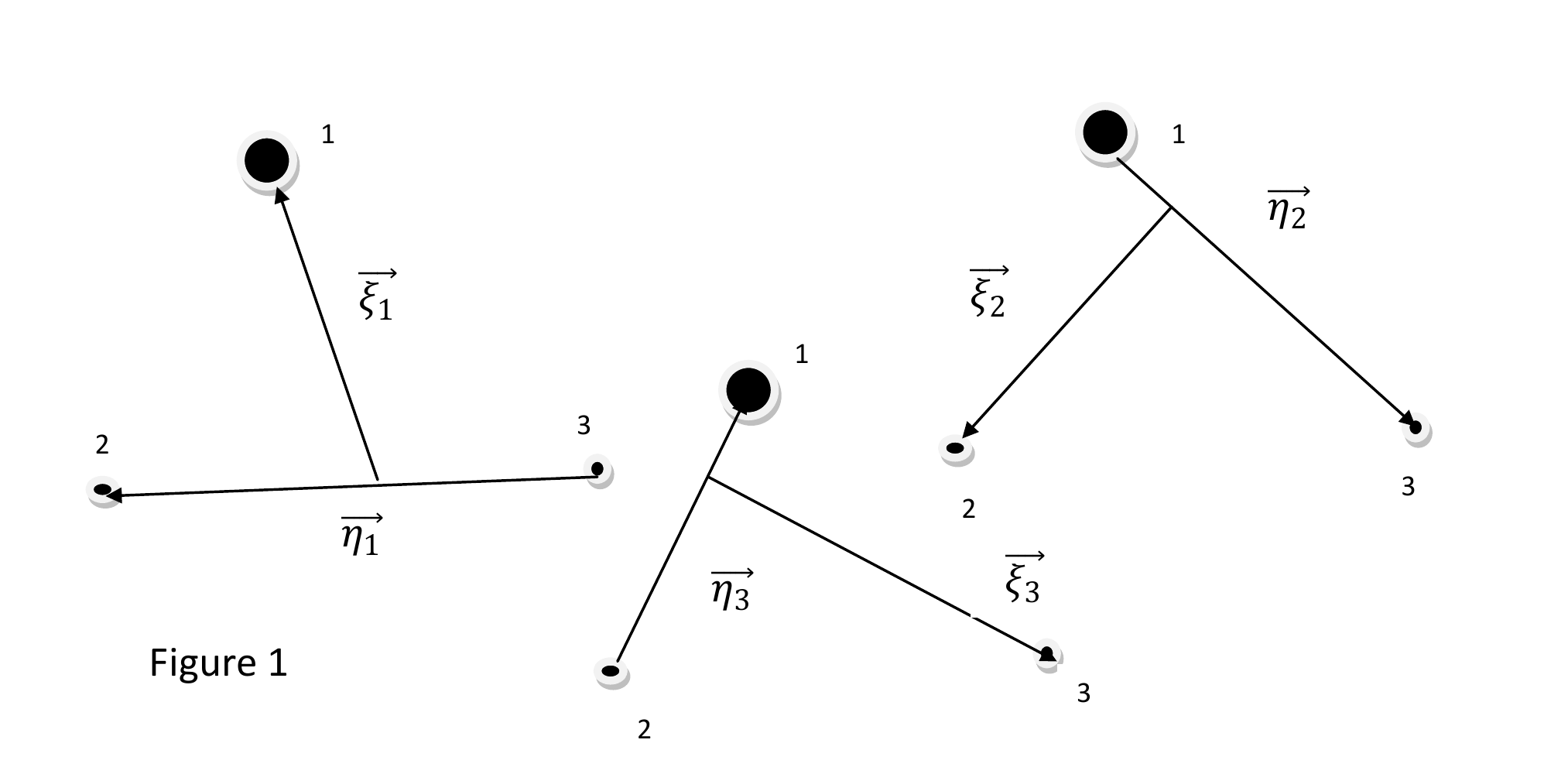}}
\caption{Choice of Jacobi coordinates in different partitions of a three-body system.}
\label{fig:boxed_graphic}
\end{figure}
The Jacobi coordinates [10] in the $i^{th}$ partition can be defined as: 
\begin{equation}
 \left.  \begin{array}{ccl}
   \vec{\eta_{i}} & = &C(\vec{r_{j}} - \vec{r_{k}}) \\ 
   \vec{\zeta_{i}} & = &C^{-1} \left( \vec{r_{i}} - \frac{m_{j}
\vec{r_{j}} + m_{k} \vec{r_{k}}}{ m_{j} + m_{k}} \right) \\
\vec{R} & = & (m_i\vec{r_i}+ m_j\vec{r_j} + m_k\vec{r_k})/M
    \end{array}  \right\} 
\end{equation}
 where $C=\left[ \frac{m_{j} m_{k}M}{m_{i}(m_{j}+m_{k})^{2}}
\right] ^{\frac{1}{4}}$, $M=m_i+m_j+m_k$ and the sign of $\vec{\eta_{i}}$ is determined by the condition that ($ i, j, k $) should form a cyclic permutation of (1, 2, 3).\\
The Jacobi coordinates are connected to the hyperspherical coordinates [11] as
\begin{equation}
 \left. \begin{array}{cclcccl}
 \eta_{i} & = & \rho \cos \phi_{i}&;& \zeta_{i} & = & \rho \sin \phi_{i}\\
\rho&=& \sqrt{ \eta_{i}^{2} +\zeta_{i}^{2}}&;&\phi_i&=&tan^{-1}(\zeta_i/\eta_i)
   \end{array} \right\} 
\end{equation}
The relative three-body Schr\H{o}dinger's equation in hyperspherical coordinates can be written as
\begin{equation}
\left[ - \frac{\hbar^{2}}{2\mu}\left\{ \frac{\partial^2}{\partial\rho^2}+ \frac{5}{\rho} \frac{\partial}{\partial\rho}+
\frac{\hat{{K}}^{2}(\Omega_{i})}{\rho^{2}} \right\}+ V (\rho, \Omega_{i}
) - E \right] 
\Psi (\rho, \Omega_{i} ) \:=\: 0
\end{equation}
where $\Omega_{i} \rightarrow \{ \phi_{i}, \theta_{\eta_{i}}, \phi_{\eta_{i}},
\theta_{\zeta_{i}}, \phi_{\zeta_{i}} \}$, effective mass ${\mu=\left[\frac{m_{i} m_{j} m_{k}}{M}\right]^{\frac{1}{2}}}$, potential $V(\rho, \Omega_{i})$ = $ V_{jk} + V_{ki} + V_{ij}$. The square of hyper angular momentum operator $\hat{K}^{2}(\Omega_{i})$ satisfies the eigenvalue equation [11]
\begin{equation}
\hat{{K}}^{2}(\Omega_{i}) {\cal Y}_{K \alpha_{i}}(\Omega_{i})=K
(K + 4 )  
{\cal Y}_{K \alpha_{i}}(\Omega_{i})
\end{equation}
where the eigen function ${\cal Y}_{K \alpha_{i}}(\Omega_{i})$ is the hyperspherical harmonics (HH). An explicit expression for the HH with specified grand orbital angular momentum $L(=\mid\vec{l_{\eta_i}} + \vec{l_{\zeta_i}}\mid)$ and its projection $M$ is given by
\begin{equation}
 \begin{array}{rcl}
{\cal Y}_{K \alpha_{i}}(\Omega_{i})& \equiv &
{\cal Y}_{K l_{\eta_{i}} l_{\zeta_{i}} L M}(\phi_{i}, \theta_{\eta_{i}},
\phi_{\eta_{i}}, \theta_{\zeta_{i}}, \phi_{\zeta_{i}})\\
 & \equiv & ^{(2)}P_{K}^{l_{\eta_{i}} l_{\zeta_{i}}}(\phi_{i})
\left[Y_{l_{\zeta_i}}^{m_{\eta_i}}(\theta_{\eta_{i}}, \phi_{\eta_{i}}) Y_{l_{\zeta_{i}}}^{
m_{\zeta_{i}}}(\theta_{\zeta_{i}}, \phi_{\zeta_{i}}) \right]_{L M}
 \end{array}
\end{equation}
with $\alpha_{i}\equiv \{l_{\eta_{i}}, l_{\zeta_{i}}, L, M\}$ and $[ ]_{L M}$ denoting angular momentum coupling.
The hyper-angular momentum quantum number $K$($=2n_i+ l_{\eta_{i}} + l_{\zeta_{i}}$; $n_i\rightarrow$ a non-negative integer) is not a conserved quantity for the three-body system.
In a chosen partition (say partition $\lq\lq i$"), the  wave-function $\Psi(\rho, \Omega_{i})$ is expanded in the complete set of HH 
\begin{equation}
\Psi(\rho, \Omega_{i}) = \sum_{K\alpha_{i}}\rho^{-5/2}U_{K\alpha_{i}}   
  (\rho) {\cal Y}_{K\alpha_{i}}(\Omega_{i})
\end{equation}
Substitution of Eq. (6) in Eq. (3) and use of orthonormality of HH, leads to a set of coupled
differential equations (CDE) in $\rho$
\begin{equation}
\begin{array}{cl}
& \left[ -\frac{\hbar^{2}}{2\mu} \frac{d^{2}}{d\rho^{2}}
+\frac{(K+3/2)(K+5/2) \hbar^2}{2\mu\rho^{2}} - E \right] 
U_{K \alpha_{i}}(\rho)  \\
+ & \sum_{K^{\prime} \alpha_{i}^{\prime}} <K \alpha_{i}
\mid V(\rho, \Omega_{i}) \mid K^{\prime} \alpha_{i}^{\prime}
>U_{K^{\prime} \alpha_{i}^{~\prime}}(\rho) \: = \: 0.
\end{array}
\end{equation}
where
\begin{equation}
<K \alpha_{i} |V(\rho,\Omega_i)| K^{\prime}, \alpha_{i}^{\prime}> = \int
{\cal Y}_{K\alpha_{i}}^{*}(\Omega_{i}) V(\rho, \Omega_{i}) {\cal
Y}_{K^{\prime} 
 \alpha_{i}^{~\prime}}(\Omega_{i}) d\Omega_{i}
\end{equation}
 
\section{Results and discussions} 
For the present calculation, we assign the label $\lq i$' to the nucleus of mass $m_i=M_N=Am_N$ ($m_N\rightarrow$ nucleon mass, A$\rightarrow$ mass number), the label $\lq j$' to the negatively charged muon of mass $m_j=m_{\mu}$ (and charge -e) and the label $\lq k$' to the electron of mass $m_k=m_e$ (and charge -e). Hence, for this particular choice of masses, Jacobi coordinates of Eq. (1) in the partition $\lq\lq i$" become 
\begin{equation}
 \left. \begin{array}{rcl}
  \vec{\eta_{i}} & = &C (\vec{r_{j}} - \vec{r_{k}}) \\
  \vec{\zeta_{i}} & = &C^{-1} (\vec{r_{i}} - \frac{m_{\mu}\vec{r_{j}}+
m_e\vec{r_{k}}}{m_{\mu}+m_e}) 
   \end{array} \right\}
\end{equation}
and the corresponding Schr\"{o}dinger equation (Eq. (7)) is  
\begin{equation}
\begin{array}{lcl}
 \left[-\frac{\hbar^2}{2\mu}\left\{ \frac{d^{2}}{d\rho^{2}}
-\frac{(K+3/2)(K+5/2)}{\rho^{2}}\right\} - E \right] 
U_{K \alpha_{i}}(\rho)&&  \\
+ \sum_{K^{\prime} \alpha_{i}^{~\prime}} <K\alpha_{i}
\mid\frac{C\exp(-\lambda\rho\cos\phi_i)}{\rho 
cos\phi_{i}} - \frac{Z\exp(-\lambda\rho\left|C \sin \phi_{i}~
\hat{\zeta_{i}}-
\frac{1}{2C} \cos\phi_{i}\hat{\eta_{i}}\right|)}{\rho\left|C \sin \phi_{i}~
\hat{\zeta_{i}}-
\frac{1}{2C} \cos\phi_{i}\hat{\eta_{i}}\right|}&& \\
  -\frac{Z\exp(-\lambda\rho\left|
C\sin \phi_{i} \hat{\zeta_{i}}+\frac{1}{2C}\cos\phi_{i}
\hat{\eta_{i}}\right|)}{\rho\left|
C\sin \phi_{i} \hat{\zeta_{i}}+\frac{1}{2C}\cos\phi_{i}
\hat{\eta_{i}}\right|}
             \mid  K^{\prime} \alpha_{i}^{~\prime}
>U_{K^{\prime} \alpha_{i}^{~\prime}}(\rho)& = & 0
\end{array}
\end{equation}
where $C=\left[\frac{m_{\mu}m_e(M_{N}+m_{\mu}+m_e)}{M_N(m_{\mu}+m_e)^2} \right]^{\frac{1}{4}}$, $\mu=\left(\frac{M_{N}m_{\mu}m_e}{M_{N}+m_{\mu}+m_e} \right)^{\frac{1}{2}}$ is the effective mass of the system and $\lambda$ ($\geq 0$) is an adjustable screening parameter. In atomic units we take $\hbar^{2}=m_e=m=e^{2}=1$. Masses of the particles involved in this work are partly taken from [11-14]. 

In the ground state $(1S^{\mu}1S^e)$ of $N^{Z+}\mu^-e^-$ three-body system, the total orbital angular momentum, $\mid \vec{L}\mid=\mid\vec{l_{\eta_{i}}}+\vec{l_{\zeta_{i}}}\mid=0$ and there is no restriction (on $l_{\eta_i}$) due to Pauli requirement as electron and muon are non-identical fermions. Since $L=0$, ${l_{\eta_{i}}= l_{\zeta_{i}}}$, and the set of quantum numbers $\alpha_{i}$ is ${\left\{ l_{\eta_{i}}, l_{\eta_{i}}, 0, 0 \right\}}$. Hence, the set of quantum numbers ${\left\{K\alpha_{i} \right\}}$ can be represented by ${\left\{K l_{\eta_{i}}\right\}}$ only. Eq. (10) is solved following the method described in our previous work [11] to get the ground state energy E.

One of the major drawbacks of HH expansion technique is its slow convergence for Coulomb-type long range potentials, unlike for the Yukawa-type short-range potentials for which the convergence is reasonably fast [10,15]. So, to achieve the desired degree of convergence, sufficiently large $K_m$ value has to be included in the calculation. But, if all $K$ values up to a maximum of $K_m$ are included in the HH expansion then the number of the basis states and hence the number of CDEs involved equals $(1+K_m/2)(1+K_m/4)$. Which shows a rapid increase in the number of basis states and hence the size of coupled differential equations (CDE) (Eq. (7)) with increase in $K_m$. For the available computer facilities, we are allowed to solve up to $K_m=28$ only, reliably. The calculated ground state energies ($B_{K_m}$) with increasing $K_m$ for some representative cases are presented in columns 2, 4 and 6 of Table I. Energies for a number of muonic atom/ions of different atomic number (Z) at $K_m=28$ are presented in column 4 of Table II. The calculated energies at $K_{m}=\infty$ of column 5 in Table II are obtained by extrapolating those obtained at $K_m\leq 28$. The pattern of convergence of the energy of the lowest bound S-state with respect to increasing $K_m$ can be checked by gradually increasing $K_m$ values in suitable steps ($dK$) and comparing the corresponding relative energy difference $\chi =\frac{B_{K+dK}-B_K}{B_{K+dK}}$ with that found in the previous step. From Table I, it can be seen that at $K_m=28$, the energy of the lowest bound S-state of ${e^-\mu^-}$ $^{\infty}$He$^{2+}$ converges up to 4th decimal places and similar convergence trends are observed in the remaining cases.

\begin{figure}
\flushleft
\fbox{\includegraphics[width=0.30\linewidth, height=0.25\linewidth]{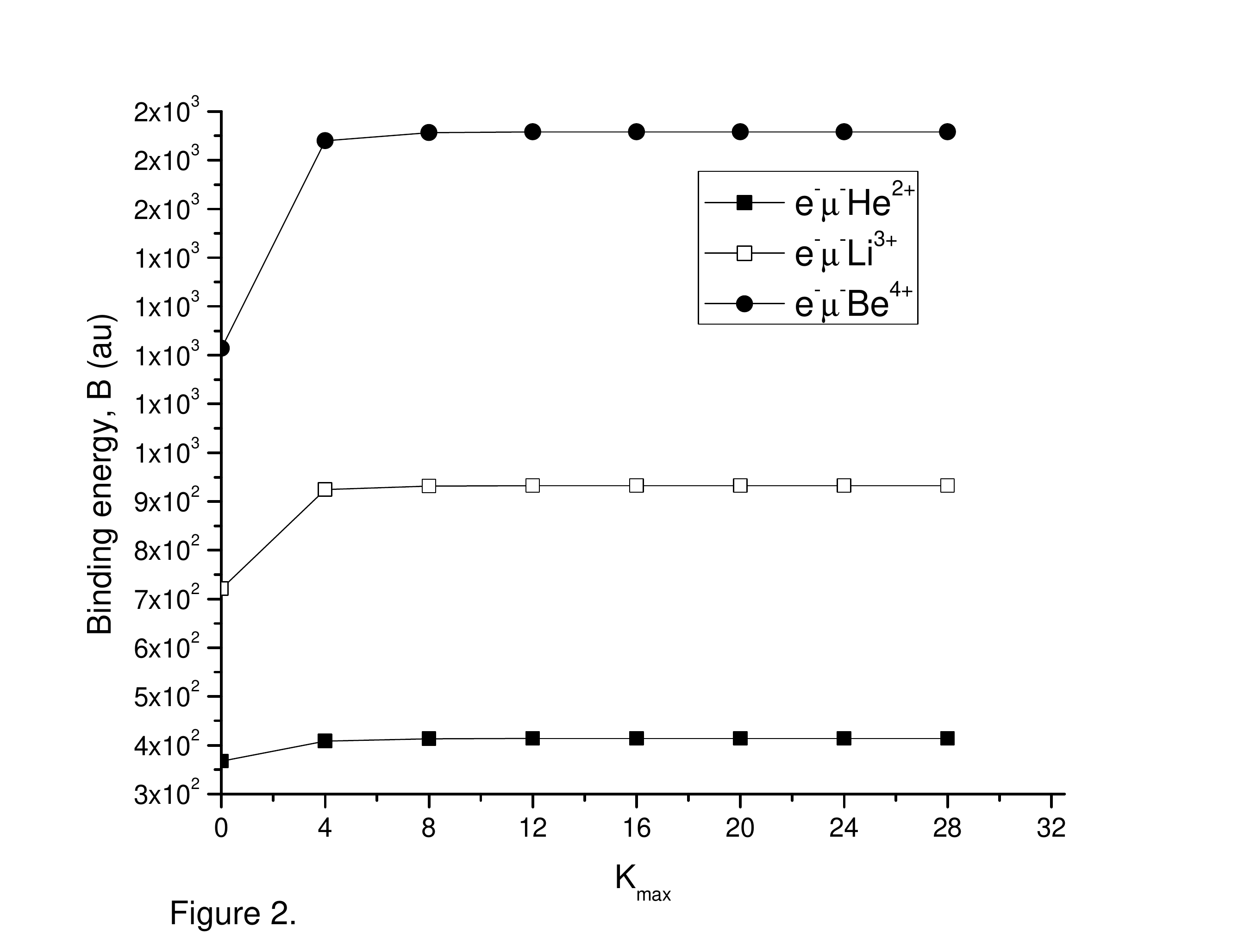}}
\fbox{\includegraphics[width=0.30\linewidth, height=0.25\linewidth]{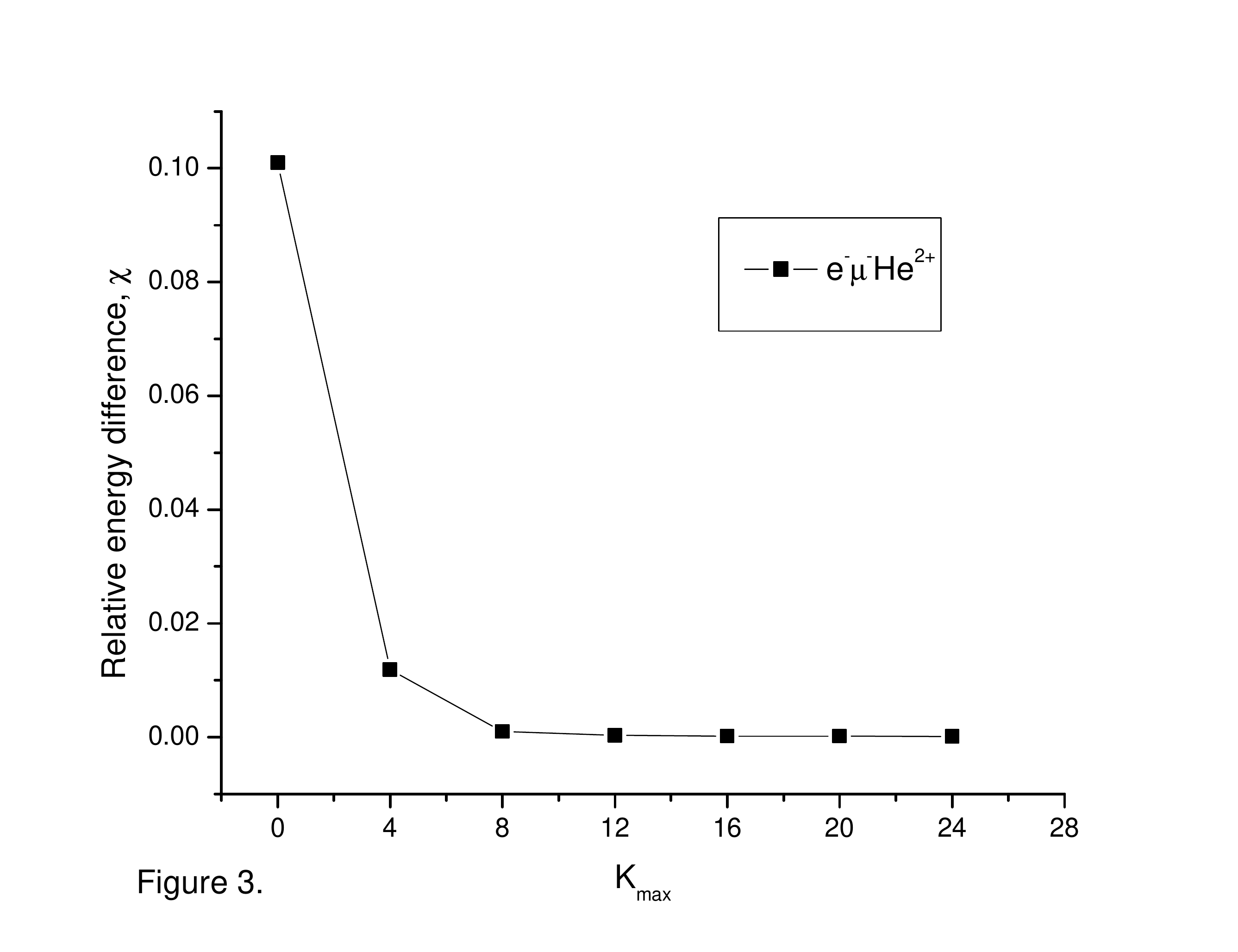}}
\fbox{\includegraphics[width=0.30\linewidth, height=0.25\linewidth]{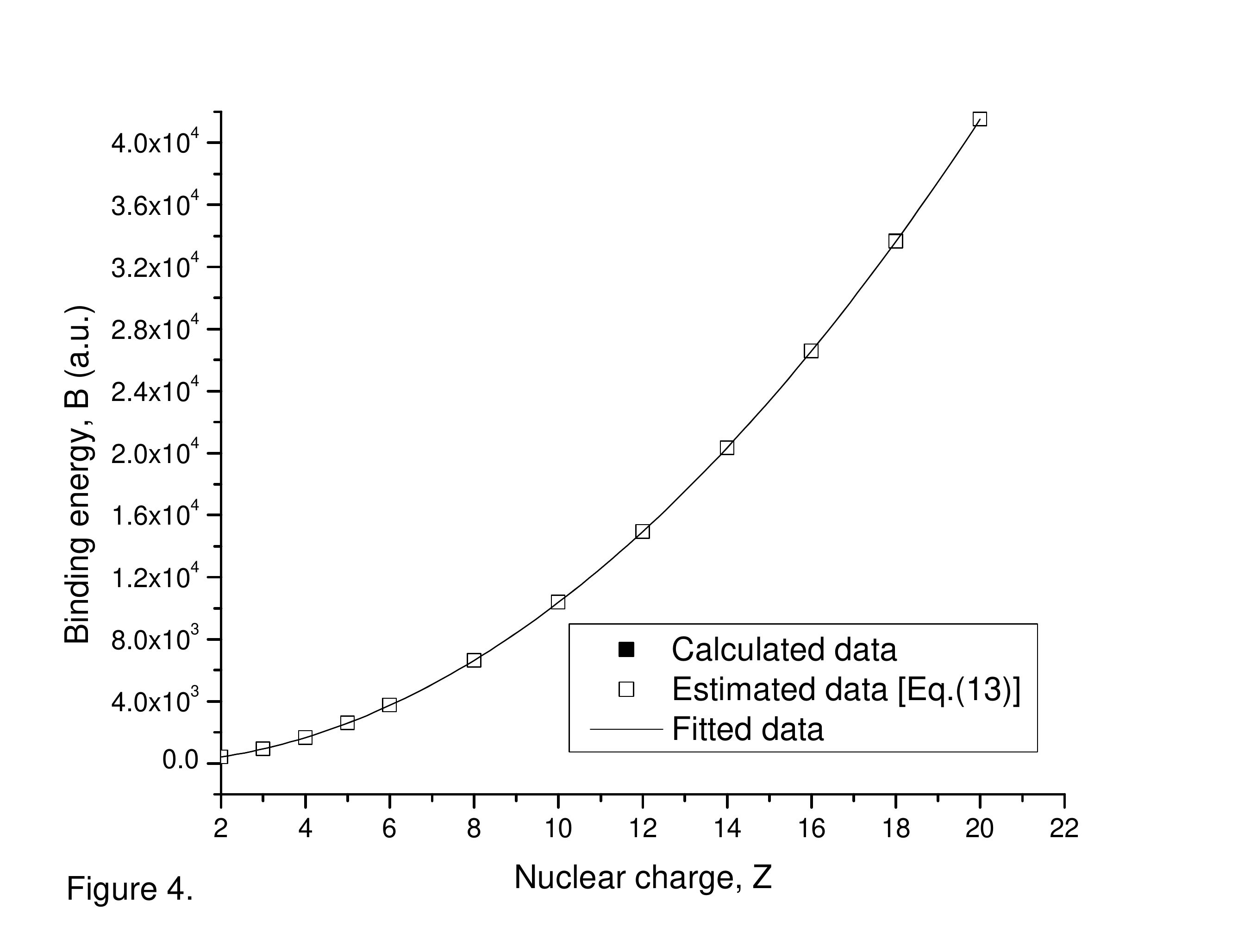}}
\caption{Pattern of dependence of the ground-state energy (B= -E) of muonic atom/ions on the increase in $K_{max}$.}
\caption{Pattern of dependence of the ground-state relative energy difference $\chi=\frac{B_{K_m+4}-B_{K_m}}{B_{K_m+4}}$ of muonic helium ($^{\infty}$He$^{2+}\mu^-e^-$) on the increase in $K_{max}$.}
\caption{Pattern of dependence of the ground-state energy (B) of muonic atom/ions on the increase in nuclear charge Z.}
\label{fig:boxed_graphic}
\end{figure}

 The pattern of increase in binding energy, B (=-E) with respect to increasing $K_{max}$ is shown in Figure 2 for few representative cases. In Figure 3 the relative energy difference $\chi$ is plotted against $K_{max}$ to demonstrate the relative convergence trend in energy. The calculated ground state energies for muonic three-body systems of infinite nuclear mass but of different nuclear charge Z, have been plotted against Z as shown in Figure 4 to study the Z-dependence of the binding energies. The curve of Figure 4 shows a gradual increase in energy with increasing Z approximately following the formula
\begin{eqnarray}    B(Z)&=&0.50137-1.00067Z+103.88149Z^2-4.2653\times 10^{-6}Z^3+8.80546\times 10^{-8}Z^4 \end{eqnarray}
Eq. (11) may be used to estimate the ground state energy of muonic atom/ions of given Z and of infinite nuclear mass. Finally, in Table II, energies of the lowest bound S-state of several muonic three-body systems obtained by numerical solution of the coupled differential equations by the renormalized Numerov method [16] have been compared with the ones of the literature wherever available. Since reference values are not available for systems having nuclear charge $Z>3$, we made a crude estimation of the ground-state ($1S^{\mu}1S^e$) energies following the relation
 \begin{equation} B_{est}^{3B} = \frac{Am_NZ^2}{2}\left[\frac{1}{1+Am_N}+\frac{m_{\mu}}{m_{\mu}+Am_N}\right]\end{equation} 
where  $m_N=1836$, muon mass $m_{\mu}=206.762828$ (in atomic unit) and A is mass number of the nucleus. Here we assumed two hydrogen-like subsystems for the muonic atom/ions. This estimate can further be improved by assuming a compact $(A(Z),\mu^-)_{1S}$ positive muonic ion and an electron in the 1S state, "feeling" a (Z-1) charge and an $Am_N+m_{\mu}$ mass. For this case the estimation formula (Eq. (12)) can be replaced by 
\begin{equation} B_{est}^{2B} = \frac{1}{2}\left[\frac{Am_N+m_{\mu}}{1+Am_N+m_{\mu}}(Z-1)^2+\frac{Am_Nm_{\mu}}{m_{\mu}+Am_N}Z^2\right]\end{equation} 
The energies estimated by Eq. (13)) is shown in column 3 of Table II.

\section{Conclusion}
A comparison among the different results tabulated in columns 2 to 4 of Table II reveals that the estimation formula given by  Eq. (13) gives better result in comparison with the highly accurate variational results listed in bold in column 4 of Table II. This indicates the fact that the three-body systems ($A(Z),\mu^-,e^-$) form weakly bound three-body states very close to the ($A(Z),\mu^-)_{1S}+e^-$ threshold. Again, the convergence trend of the three-body wave function and energy for the muonic systems for the long range Coulomb potential is poor even at $K_{max}=28$. It can be seen from Table II, that, between the estimated energies $B^{3B}_{est}$ in column 2 and $B^{2B}_{est}$ in column 3, the latter are closer to the calculated energies in columns 4 and 5. This indicates that the muonic three-body systems have more possibility of having super heavy-hydrogen-like two-body structure over three helium-like three-body structure. Finally, it can be noted that a further improvement in the calculations could have been done by incorporating the Kato's cusp conditions[17] in the limit $r_{jk}\rightarrow 0$ appropriately.\\
\hspace*{1cm} The authors acknowledge Aliah University for providing computer facilities and one of the authors M. Hasan gladly acknowledges Aliah University for granting a research assistantship.

\section{Figure Caption}
\begin{list}{}{}
\item Fig. 1. Choice of Jacobi coordinates in different partitions of a three-body system.
\item Fig. 2. Pattern of dependence of the ground-state energy, B (=-E) of muonic atom/ions on the increase in $K_{max}$.
\item Fig. 3. Pattern dependence of the ground-state relative energy difference $\chi=\frac{B_{K_m+4}-B_{K_m}}{B_{K_m+4}}$ of muonic helium ($^{\infty}$He$^{2+}\mu^-e^-$) on the increase in $K_{max}$.
\item Fig. 4. Pattern of dependence of the ground-state energy (B) of muonic atom/ions on the increase in nuclear charge Z.
\end{list}

\section{Tables}
\begin{table}
\begin{center}
\begin{scriptsize}
{\bf Table I. Convergence trend in the lowest s-state energy (-E) with increasing $K_{max}$ in representative cases of muonic three-body systems.}\\
\vspace{5pt}
\begin{tabular}{ccccccc}\hline
&\multicolumn{6}{c}{Energy (-E) and the corresponding relative energy difference ($\chi$)}\\
\cline{2-7}
System$\rightarrow$&\multicolumn{2}{c}{$^{\infty}$He$^{2+}\mu^-e^-$}&\multicolumn{2}{c}{$^{\infty}$Li$^{3+}\mu^-e^-$}&\multicolumn{2}{c}{$^{\infty}$Be$^{4+}\mu^-e^-$}\\
\cline{2-3}\cline{4-5}\cline{6-7}
$K_{max}$&$-E$&$\chi$&$-E$&$\chi$&$-E$&$\chi$\\\hline
0&367.23937&0.100954&720.76709&0. 220429&1214.19122&0.259639\\
4&408.47659&0.011810&924.57004&0.007488& 1640.17071&0.009872\\
8&413.35837&0.000962&931.54539&0.000710&1656.35200&0.000842\\
12&413.75620&0.000316&932.20649&0.000113&1657.74786&0.000125\\
16&413.88690&0.000146&932.31191&0.000071&1657.95455&0.000071\\
20&413.94748&0.000143&932.37854&0.000054&1658.07252&0.000030\\
24&414.00669&0.000075&932.42892&0.000022&1658.12307&0.000006\\
28&414.03768& &932.44956& &1658.13278&\\\hline
\end{tabular}
\end{scriptsize}
\end{center}
\end{table}
\begin{table}
\begin{center}
\begin{scriptsize}
{\bf Table II. Energy (B) of the lowest bound S-state of electron-muon-nucleus three-body systems.}\\
\vspace{5pt}
\begin{tabular}{lrrrrl}\hline
System&\multicolumn{4}{c}{Binding energies expressed in atomic unit (a.u.)}\\
\cline{2-6}
& \multicolumn{2}{c}{Estimated}&\multicolumn{2}{c}{Calculated}&{Screening }\\
\cline{2-5}
&$B_{est}^{3B}[Eq. (12)]$&$B_{est}^{2B}[Eq. (13)]$& $B^{calc}_{K_m=28}$&$B^{calc}_{K_m=\infty}$ & parameter,$\lambda$\\\hline
$e^-\mu^-$$^3$He$^{2+}$&400.574&399.064&399.042&399.085&56.12614\\
&&&&{\bf 399.042$^a$, 399.043$^b$}&\\
$e^-\mu^-$$^4$He$^{2+}$&404.212&402.702&402.637&402.679&55.9957\\
&&&&{\bf 402.637$^c$, 402.641$^d$}&\\
$e^-\mu^-$$^{\infty}$He$^{2+}$&415.537&414.026&414.038&414.078&55.6172\\
&&&&{\bf 414.036$^e$, 414.037$^f$}&\\
$e^-\mu^-$$^6$Li$^{3+}$&917.814&915.291&915.231&915.263&54.49577\\
&&&&{\bf 915.231$^e$, 915.231$^g$}&\\
$e^-\mu^-$$^7$Li$^{3+}$&920.224&917.701&917.650&917.682&54.47998\\
&&&&{\bf 917.649$^e$, 917.650$^g$}&\\
$e^-\mu^-$$^{\infty}$Li$^{3+}$&934.957&932.433&932.449&932.481&54.38531\\
&&&&{\bf 932.457$^e$}&\\
$e^-\mu^-$$^{9}$Be$^{4+}$& 1641.703&1638.161&1638.121 &1638.159&55.14367 \\
$e^-\mu^-$$^{\infty}$Be$^{4+}$&1662.146&1658.603&1658.133&1658.156&55.091\\

$e^-\mu^-$$^{10}$B$^{5+}$ &2568.319&2563.753&2563.765&2563.799&32.3766\\
$e^-\mu^-$$^{\infty}$B$^{5+}$&2597.104 &2592.535&2591.810&2592.073 &35.356\\

$e^-\mu^-$$^{12}$C$^{6+}$&3705.224 &3699.628&3699.604 &3699.725&57.4880\\ 
$e^-\mu^-$$^{\infty}$C$^{6+}$&3739.829&3734.231&3734.342&3734.419 &57.47\\

$e^-\mu^-$$^{16}$O$^{8+}$&6602.337&6594.666&6594.649&6594.915&48.4763\\ 
$e^-\mu^-$$^{\infty}$O$^{8+}$&6648.585&6640.910&6640.997&6641.169 &60.086\\

$e^-\mu^-$$^{20}$Ne$^{10+}$&10330.524 &10320.754&10320.754 &10321.194&54.840675\\ 
$e^-\mu^-$$^{\infty}$Ne$^{10+}$&10388.414&10378.641&10377.689&10378.133 &54.895\\										
$e^-\mu^-$$^{24}$Mg$^{12+}$&14889.783 &14877.894&14877.894 &14878.527&58.432895\\
$e^-\mu^-$$^{\infty}$Mg$^{12+}$&14959.316&14947.424&14947.838&14948.475&58.461\\

$e^-\mu^-$$^{28}$Si$^{14+}$&20280.115 &20266.085&20266.084 &20266.796&66.473555\\
$e^-\mu^-$$^{\infty}$Si$^{14+}$&20361.291&20347.257&20348.161&20348.876&66.472\\

$e^-\mu^-$$^{32}$S$^{16+}$&26501.520&26485.328&26485.331&26487.764&75.71491\\
$e^-\mu^-$$^{\infty}$S$^{16+}$&26594.340&26578.142&26578.177&26579.097&68.1302\\

\footnote{•} $e^-\mu^-$$^{40}$Ar$^{18+}$&33564.416&33546.038&33546.191&33548.211&78.31812\\
$e^-\mu^-$$^{\infty}$Ar$^{18+}$&33658.461&33640.078&33662.298&33665.494&78.3021\\

$e^-\mu^-$$^{40}$Ca$^{20+}$&41437.550&41416.966&41416.889&41418.005&66.5045\\
$e^-\mu^-$$^{\infty}$Ca$^{20+}$&41553.656&41533.066&41534.805&41535.924&66.509\\\hline
\end{tabular}\\
$^a$Ref[2,18-20,26], $^b$Ref[21], $^c$Ref[2,18-19,21-22], $^d$Ref[1,23-24], 
$^e$Ref[2], $^f$Ref[19], $^g$Ref[25]\\
\end{scriptsize}
\end{center}
\end{table}
\end{document}